# Wavelength conversion through plasmon-coupled surface states

Deniz Turan[1], Ping Keng Lu[1], Nezih T. Yardimci[1], Zhaoyu Liu[2], Liang Luo[2], Joong-Mok Park[2], Uttam Nandi[3], Jigang Wang[2], Sascha Preu[3], and Mona Jarrahi[1,*]

[1]Electrical and Computer Engineering Department, University of California, Los Angeles, CA, 90095, USA.

[2]Department of Physics and Astronomy and Ames Laboratory-U.S. DOE, Iowa State University, Ames, Iowa 50011, USA.

[3]Department of Electrical Engineering and Information Technology, Technical University Darmstadt, 64283 Darmstadt, Germany.

*Corresponding author. Email: mjarrahi@ucla.edu

**Abstract:** Surface states generally degrade semiconductor device performance by raising the charge injection barrier height, introducing localized trap states, inducing surface leakage current, and altering the electric potential. Therefore, there has been an endless effort to use various surface passivation treatments to suppress the undesirable impacts of the surface states (*1–5*). We show that the giant built-in electric field created by the surface states can be harnessed to enable passive wavelength conversion without utilizing any nonlinear optical phenomena. Photo-excited surface plasmons are coupled to the surface states to generate an electron gas, which is routed to a nanoantenna array through the giant electric field created by the surface states. The induced current on the nanoantennas, which contains mixing product of different optical frequency components, generates radiation at the beat frequencies of the incident photons. We utilize the unprecedented functionalities of plasmon-coupled surface states to demonstrate passive wavelength conversion of nanojoule optical pulses at a 1550 nm center wavelength to terahertz regime with record-high efficiencies that exceed nonlinear optical methods by 4-orders of magnitude. The presented scheme can be used for optical wavelength conversion to different parts of the electromagnetic spectrum ranging from microwave to infrared regimes by using appropriate optical beat frequencies.

**Main Text:** When a semiconductor lattice is terminated on the surface, the periodicity of the lattice is broken since the surface atoms do not have sufficient number of atoms that they can bond to, leaving behind incomplete chemical bonds. These so called dangling bonds produce localized surface states with energy levels that are located within the bandgap of the semiconductor (*6–9*). The Fermi energy level at the surface of a semiconductor is fixed to the energy level at which the surface state density peaks, while the Fermi energy level away from the semiconductor surface is determined by the semiconductor doping. Therefore, the presence of the surface states takes away a very important degree of freedom for engineering semiconductor devices by altering the electric potential profile and is generally a major source of degradation in semiconductor devices.

Despite the endless efforts to suppress the surface semiconductor states, they have unique electrochemical properties that are not provided by bulk semiconductors and could enable unprecedented device functionalities. Figure 1A illustrates how the presence of surface states induces a giant built-in electric field at the surface of a p-doped InAs semiconductor, which exceeds the breakdown field of bulk InAs. The energy level at which the surface state density of InAs peaks is located above its direct bandgap because there is a large difference between the direct and indirect bandgap energies of InAs (*7, 8*). Since the electrons that occupy the surface states have an average total energy higher than the bulk InAs, they migrate from the surface states to the bulk InAs to reach equilibrium, leaving behind immobile charge of uncompensated donor ions, which produces a giant built-in electric field.

To effectively utilize this built-in electric field for optical wavelength conversion, optical photons excite a nanoantenna array to couple photo-excited surface plasmons to the surface states (Fig. 1B). Excitation of surface plasmons enhances the optical intensity and photoabsorption near the InAs surface (*10-22*), where the strength of the built-in electric field is maximized. The absorbed photons generate a tightly confined electron gas under the nanoantenna contacts with an electron concentration that resonates at the mixing product of different optical frequency components. This electron gas swiftly drifts to the nanoantennas through the built-in electric field. The induced current on the nanoantennas generates radiation at the beat frequencies of the optical photons. Figure 2A shows a nanoantenna array designed to couple photo-excited surface plasmons to the InAs surface states where a built-in electric field drifts the photo-induced electron gas to the nanoantennas to generate radiation at the optical beat frequencies. Unlike the bulky and complex nonlinear optical setups that require high-energy lasers, tight optical focus, and/or tilted beam to provide high optical pump intensity and phase matching for efficient wavelength conversion, wavelength conversion through plasmon-coupled surface states does not require a complex optical setup and is not sensitive to optical focus and alignment (see supplementary materials). Figure 2B shows a fabricated nanoantenna array on InAs that is simply glued at the tip of an optical fiber without using any intermediate optical component and can be pumped by a compact fiber laser. We experimentally demonstrate conversion of 3.68 nanojoule optical pulses with a 150 fs pulsewidth coupled to the fiber at a 1550 nm center wavelength to 1.78 picojoule terahertz pulses radiated from the nanoantenna array (see supplementary materials) with more than a 4 THz bandwidth and 105 dB dynamic range (Fig. 2C). Broader radiation bandwidths exceeding 6 THz and higher dynamic ranges exceeding 110 dB are achieved when using optical pulses with shorter pulsewidth and higher power (see supplementary materials). The measured terahertz pulse energy/power from the fabricated nanoantenna array as a function of the optical pulse energy/power (Fig. 2D inset) is compared with other passive optical-to-terahertz converters reported in the literature, which utilize nonlinear optical processes (*23-43*), spintronics (*44–47*), and the photodember effect (*48–51*). The comparison indicates record-high efficiency of the plasmon-coupled surface states in passive wavelength conversion of nanojoule optical pulses to terahertz regime with efficiencies that exceed nonlinear optical methods by 4-orders of magnitude (Fig. 2D).

In order to achieve high wavelength conversion efficiencies, the semiconductor structure and nanoantenna geometry are chosen to maximize the spatial overlap between the built-in electric field and photoabsorption profiles. The strength and extent of the built-in electric field below the InAs surface can be controlled by the doping profile of the InAs substrate. As illustrated in Fig. 3A, since the Fermi energy level at the surface of InAs is pinned above the conduction band minimum, increasing the p-type doping of the bulk results in a steeper band bending and, therefore, a stronger built-in electric field near the InAs surface. To better show the impact of the substrate doping, identical nanoantenna arrays are fabricated on three InAs substrates with p-type doping concentrations of $10^{17}$, $10^{18}$, and $10^{19}$ cm$^{-3}$ and their optical-to-terahertz conversion performance is characterized under the same optical pump beam. As predicted by the energy band diagrams illustrated in Fig. 3A, the nanoantenna array fabricated on the InAs substrate with a p-type doping concentration of $10^{19}$ cm$^{-3}$ offers the highest wavelength conversion efficiency among the three as it benefits from the highest built-in electric field near the InAs surface (Fig. 3B).

However, increasing the p-type doping reduces the extent of the built-in electric field below the InAs surface and lowers the spatial overlap between the built-in electric field and photoabsorption profiles. One way to extend the built-in electric field below the InAs surface is incorporating an undoped InAs layer between the p-doped InAs epilayer and the nanoantenna contact. As illustrated in Fig. 3C, increasing the thickness of the undoped InAs layer further extends the band bending below the InAs surface while reducing the band bending slope, indicating a tradeoff between the strength and extent of the built-in electric field in the substrate. To better show the impact of this tradeoff, identical nanoantenna arrays are fabricated on four InAs substrates with undoped InAs layer thicknesses of 0, 100, 200, and 350 nm grown on an InAs epilayer with a p-type doping of $10^{19}$ cm$^{-3}$ and their optical-to-terahertz conversion performance is characterized under the same optical pump beam. As demonstrated in Fig. 3D, the use of a 100-nm-thick undoped InAs layer increases the wavelength conversion efficiency by extending the built-in electric field in the semiconductor and increasing its spatial overlap with the photoabsorption profile. However, further increase in the thickness of the undoped InAs layer lowers the wavelength conversion efficiency due to the reduction in the built-in electric field strength. Because of its high wavelength conversion efficiency, the nanoantenna array fabricated on a 100-nm-thick undoped InAs layer grown on an InAs epilayer with a p-type doping of $10^{19}$ cm$^{-3}$ is used to demonstrate the results shown in Fig. 2.

Periodicity of the nanoantennas in the y-direction is chosen as 440 nm to provide the necessary momentum to couple the photo-excited surface plasmon waves to the interface between the metal contact and InAs substrate when excited by a TM-polarized optical beam at a 1550 nm wavelength (see supplementary materials). A 240-nm-thick $Si_3N_4$ anti-reflection coating, a 360-nm-thick nanoantenna width, and a 3/97-nm-thick Ti/Au nanoantenna height are used to increase the coupling efficiency of surface plasmon waves. Geometry of the nanoantenna array is chosen to provide high-efficiency radiation over a broad terahertz frequency range when fed with the injected electrons from the InAs substrate. Radiation power is calculated from the induced current on the nanoantennas (52). A finite-element-method-based electromagnetic solver (ANSYS-HFSS) is used to compute the induced current on nanoantennas for various geometrical parameters as a function of frequency. Figures 4A-F (top) show the induced current on the nanoantennas for different nanoantenna lengths ($L_a$) varying between 1 μm and 9 μm as a function of frequency. The steady reduction in the current amplitudes at higher frequencies is due to the non-zero transit time of the photogenerated electrons in InAs to the nanoantennas, which determines the photocurrent impulse response (see supplementary materials). Figures 4A-F (bottom) show the decomposition of the total induced current on the nanoantennas (teal lines) to the individual contributions of the injected currents from different positions of the nanoantennas (white lines) at 0.2 THz. The background color maps show the electron generation profiles computed using a Finite-difference time-domain electromagnetic solver (Lumerical), averaged over the nanoantenna width. As expected, the induced current at different nanoantenna location is proportional to the electron generation rate, which causes the ripples observed in the total induced currents. The current that is injected near the nanoantenna tip and the nanoantenna-ground line intersection has the highest contribution to the total induced current on the nanoantennas. As the injection point is moved from these margins, the induced current splits into two near-equal current components that are 180 degrees out-of-phase from one another, resulting in a destructive radiation from these out-of-phase current components (see supplementary materials). As the nanoantenna length is decreased from 9 μm (Fig. 4F) to 2 μm (Fig. 4B), the regions on the nanoantenna that do not contribute to the radiation are eliminated and the current density on the nanoantennas is increased, resulting in higher radiation powers. When the antenna length is reduced below 2 μm, the injected current to the nanoantenna is reduced because the ground lines shadow a major fraction of the optical beam, reducing the number of the photogenerated electrons in InAs (Fig. 4A). Apart from the nanoantenna length, the ground line width ($L_b$) and the gap between the nanoantenna array rows ($L_g$) also have a significant impact on the radiation efficiency. Increasing the width of the ground lines provides a lower impedance ground path for the current flow through the nanoantennas, resulting in an increase in the induced current (see supplementary materials). However, increasing the ground line width beyond 2 μm reduces the injected current to the nanoantenna because the ground lines shadow a major fraction of the optical beam, reducing the number of the photogenerated electrons in InAs. Additionally, since the photogenerated electrons inside the gap between the nanoantenna array rows do not contribute to the radiation, this gap should be kept very small to maximize the fill factor of the radiating elements. To better show the impact of the nanoantenna geometry, nanoantenna arrays with different nanoantenna lengths, ground line widths, and gaps sizes between the nanoantenna rows are fabricated with a total area of 1×1 mm$^2$ and their radiation power is characterized under the same femtosecond optical pulse illumination. As illustrated in Figs. 4H-J, the measured terahertz radiation powers are in agreement with the theoretical predictions based on the induced current profiles on the nanoantennas.

The presented wavelength conversion scheme via plasmon-coupled surface states can be used for optical wavelength conversion to different parts of the electromagnetic spectrum ranging from microwave to infrared regimes in both pulsed and continuous wave operation (see supplementary materials). Wavelength conversion efficiency can be further enhanced by boosting the built-in electric field at the semiconductor surface and increasing the spatial overlap between the built-in electric field and photoabsorption profiles. Using alternative semiconductors with a larger number of surface states above the conduction band, introducing higher p-type doping levels, and incorporating a gradient composition semiconductor ($In_{1-x}Ga_xAs$ with x increasing as a function of depth in the substrate) would introduce a steeper band-bending at the semiconductor surface and, therefore, would further enhance the built-in electric field. In addition, by growing the semiconductor active layer on a distributed Bragg reflector and an appropriate choice of nanoantenna geometry, most of the excited surface plasmons would be trapped in the semiconductor active layer and, therefore, a much stronger spatial overlap between the built-in electric field and photoabsorption profiles can be achieved.

**Acknowledgments:** We acknowledge contributions of Dr. Baolai Liang and Dr. Yuwei Fan from California Nanosystems Institute for the MBE growth of the substrates and electron beam lithography of the nanostructures, respectively. We also gratefully acknowledge the financial support of the Office of Naval Research (grant # N000141912052) and the Burroughs Wellcome Fund Innovation in Regulatory Science Award Program. Deniz Turan was supported by the Department of Energy (grant # DE-SC0016925).

**Author contributions:** DT and MJ conceived the research; DT, PKL, NTY, ZL, LL, JMP, UN, JW, and SP contributed to the experiments; DT and PKL processed the data; DT and MJ prepared the manuscript; MJ initiated and supervised the research.

**Competing interests:** The authors declare no competing financial interest.

**Data and materials availability:** All the data and methods are present in the main text and the supplementary materials.

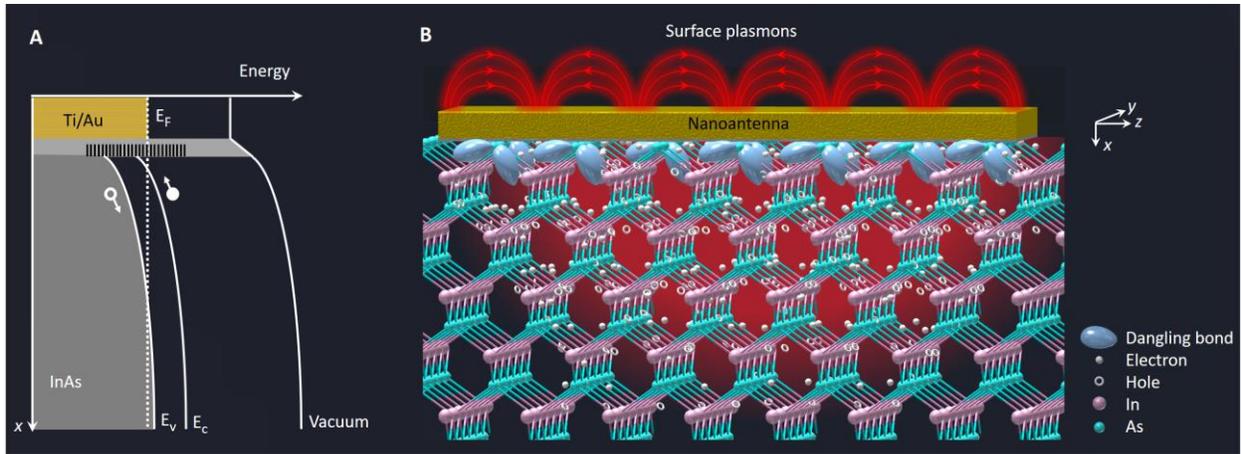

**Fig. 1. Energy band bending caused by the InAs surface states.** (**A**) Energy band diagram of a highly p-doped InAs substrate in contact with Ti/Au. The energy level at which the surface state density of InAs peaks is located above its direct bandgap because there is a large difference between the direct (0.36 eV) and indirect bandgap energies (1.21 eV) of InAs. Electrons in these surface states recombine with the holes in the valence band and occupy a part of the conduction band to minimize their total energy. As a result, the Fermi energy level ($E_F$) is pinned above the conduction band minimum ($E_c$). Free electrons in the conduction band then migrate to the p-doped InAs layer to minimize their energy further, resulting in a steep band bending and a giant built-in electric field induced at the InAs surface. (**B**) Schematic of the InAs lattice in contact with a nanoantenna that couples photo-excited surface plasmons to the surface states.

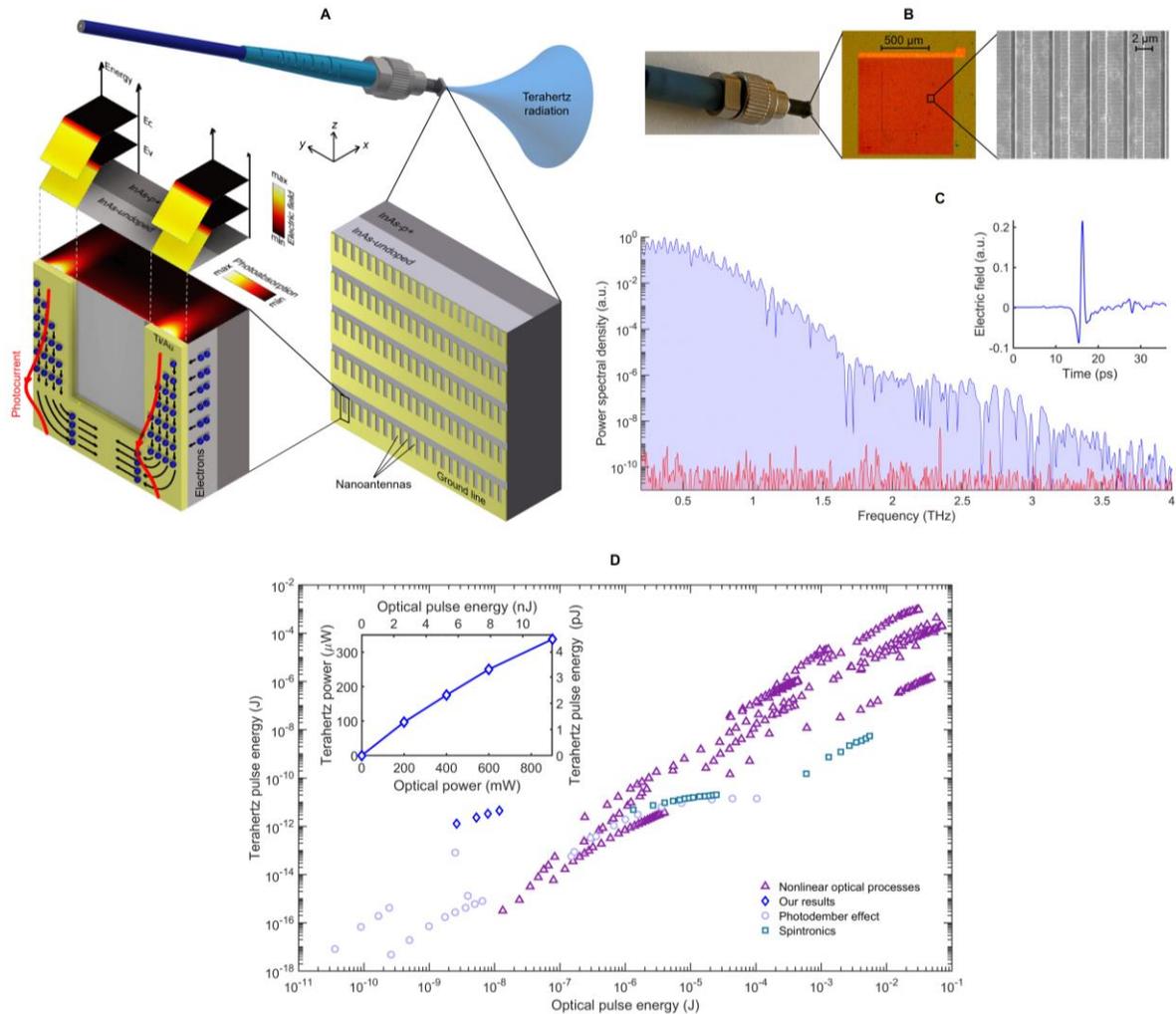

**Fig. 2. Wavelength conversion through plasmon-coupled surface states.** (**A**) Schematic of a nanoantenna array on an InAs semiconductor substrate, which is designed to couple photo-excited surface plasmons to the surface states where a built-in electric field drifts the photo-induced electron gas to the nanoantennas to generate radiation at the optical beat frequencies. The nanoantenna geometry and semiconductor structure are chosen to maximize the spatial overlap between the built-in electric field and photoabsorption profiles. (**B**) Photograph, microscopy, and scanning electron microscopy images of a fabricated nanoantenna array on a substrate consisting of a 100-nm-thick undoped InAs layer grown on a 500-nm-thick InAs epilayer with a p-type doping of $10^{19}$ cm$^{-3}$ grown on a semi-insulating GaAs substrate. (**C**) Measured terahertz radiation (in blue) and noise (in red) spectra generated from the nanoantenna array when pumped by 3.68 nanojoule optical pulses at a 1550 nm center wavelength. The time-domain radiated terahertz pulse is shown in the inset. 3200 time-domain traces are captured and averaged to resolve this spectrum. (**D**) Measured terahertz pulse energy/power from the fabricated nanoantenna array as a function of the optical pulse energy/power (inset) in comparison with other passive optical-to-terahertz converters reported in the literature.

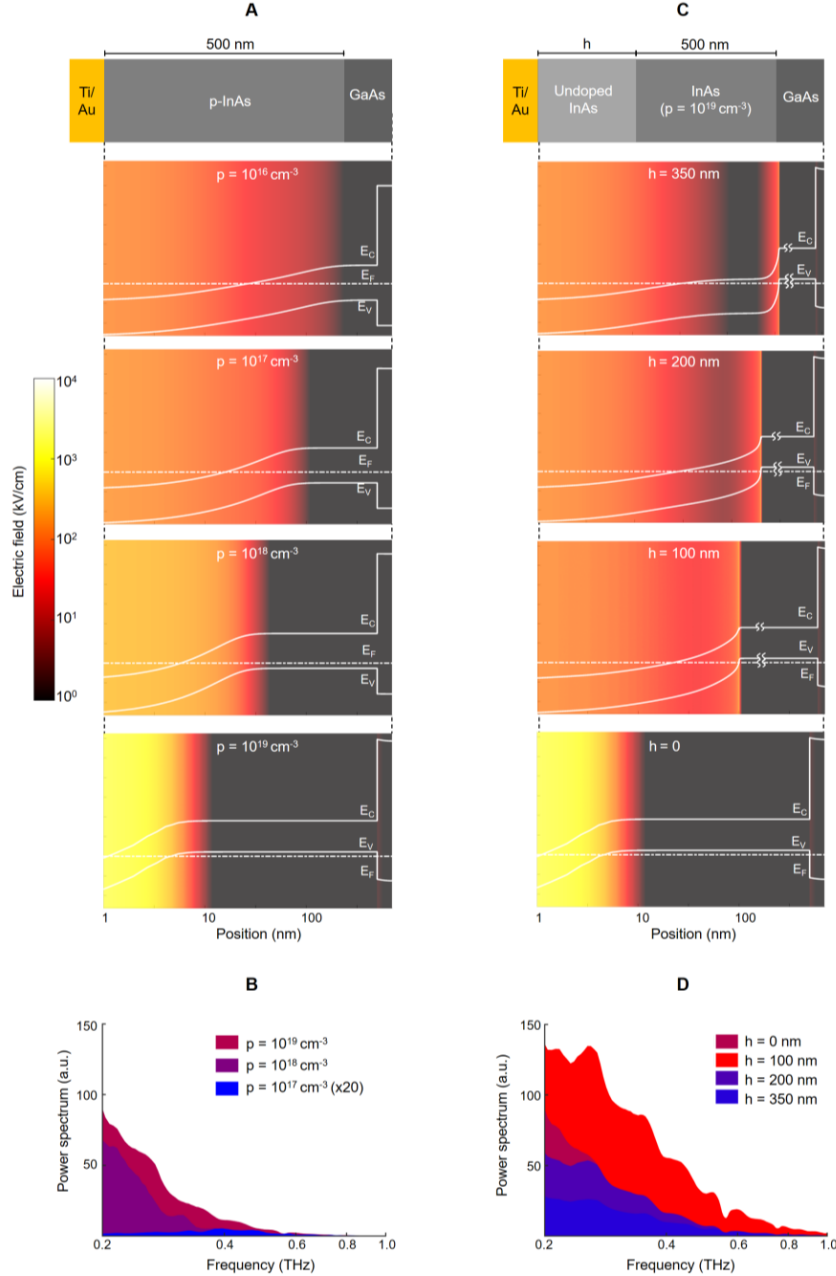

**Fig. 3. Built-in electric field profile and its impact on the wavelength conversion efficiency.** (**A**) Band diagram of the p-doped InAs layer below the Ti/Au nanoantenna contact at different p-type doping concentrations are shown in white. The color map shows the strength of the built-in electric field. Sentaurus device simulator is used to generate the band diagram and built-in electric field plots. The built-in electric field drifts the high-mobility photo-generated electrons to the Ti/Au contact without any barrier height and sweeps away the low-mobility photo-generated holes from the Ti/Au contact. (**B**) The measured terahertz radiation spectra from identical nanoantenna arrays fabricated on three InAs substrates with p-type doping concentrations of $10^{17}$, $10^{18}$, and $10^{19}$ cm$^{-3}$ in response to the same optical pump beam. The radiation spectra are shown in a linear scale to clearly show the wavelength conversion efficiency variations. (**C**) Band diagram and the built-in electric field profiles when an undoped InAs layer is incorporated between the p-doped InAs epilayer and the Ti/Au contact. (**D**) The measured terahertz radiation spectra from identical nanoantenna arrays fabricated on four InAs substrates with undoped InAs layer thicknesses of 0, 100, 200, and 350 nm grown on an InAs epilayer with a p-type doping of $10^{19}$ cm$^{-3}$.

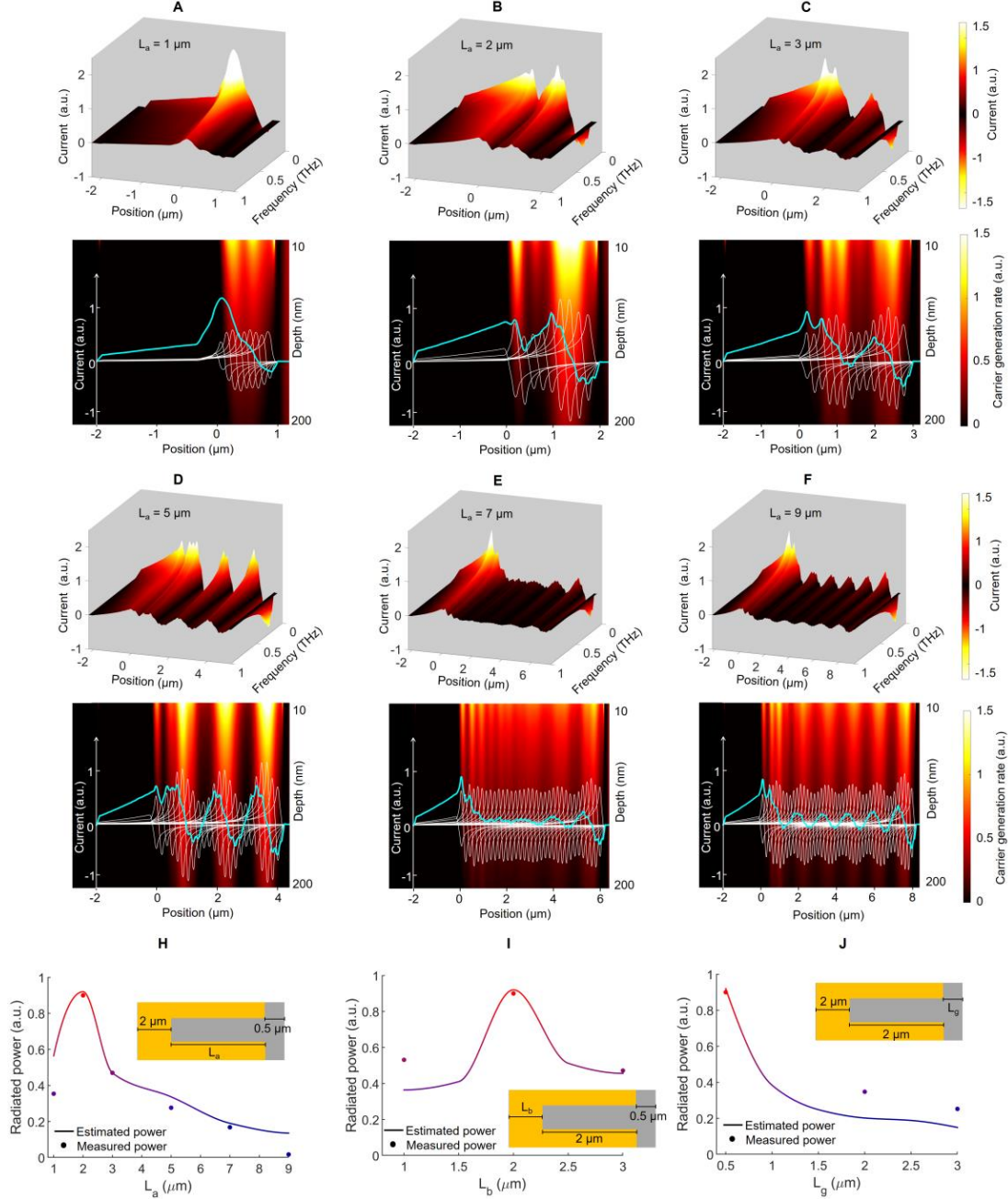

**Fig. 4. Impact of the nanoantenna geometry on the wavelength conversion efficiency:** (**A**)-(**F**) Top: The induced current on the nanoantennas as a function of frequency when the nanoantenna length ($L_a$) is varied from 1 μm to 9 μm. The ground line width ($L_b$) and the gap between the nanoantenna array rows ($L_g$) are chosen as 2 μm and 0.5 μm. The ground line is located between -2 μm and 0 positions and the nanoantenna is located between 0 and 1-9 μm positions along the z-axis. (**A**)-(**F**) Bottom: Decomposition of the total induced current on the nanoantennas (teal lines) to the individual contributions of the injected currents from different positions of the nanoantennas (white lines) at 0.2 THz. The background color maps show the electron generation profiles averaged over the nanoantenna width. Predicted and measured terahertz radiation power from fabricated nanoantenna arrays with different nanoantenna lengths, ground line widths, and gaps sizes between the nanoantenna rows are shown in (**H**), (**I**), and (**J**) respectively. All the fabricated nanoantenna arrays have a 1×1 mm² area and are characterized using the same optical pulses with 120 fs pulse width, 2.63 nJ pulse energy, and 76 MHz repetition rate.

# Supplementary Materials

**Materials and Methods**

Semiconductor Growth:

InAs layers are grown by molecular beam epitaxy (Veeco GEN-930) on semi-insulating GaAs (001) substrates. The growth is performed in an As-rich chamber at 400 °C. Be is used to dope the InAs to achieve a p-type doping concentration of $10^{19}$ cm$^{-3}$.

Device Fabrication:

The nanoantenna arrays are first defined using electron-beam lithography (Vistec EBPG 5000+ES) followed by 3/97 nm Ti/Au evaporation (CHA solution electron beam evaporator) and lift-off. Ground lines are defined by electron-beam lithography followed by a 40/360 nm Ti/Au evaporation and lift-off. Finally, a 240-nm-thick Si$_3$N$_4$ anti-reflection coating is deposited using plasma enhanced chemical vapor deposition (STS Multiplex CVD).

Radiation Measurements:

An optical parametric oscillator (OPO) pumped by a Ti:sapphire laser is used to pump the fabricated nanoantenna arrays. It provides optical pulses at a 1550 nm central wavelength, 76 MHz repetition rate and 120 fs pulsewidth. To demonstrate the fiber-coupled wavelength conversion performance shown in Fig. 2, the optical beam from the OPO is coupled to a polarization maintaining optical fiber (Thorlabs PM1550-XP) and the nanoantenna array is glued at the tip of the fiber. The optical pulsewidth incident on the nanoantenna array is increased to 150 fs due to the fiber dispersion. Terahertz radiation power measurements are performed using a calibrated pyroelectric detector (Sensors und Lasertechnik THz-30 detector calibrated by Physikalisch-Technische Bundesanstalt, Germany). Terahertz radiation spectrum measurements are performed using a terahertz time-domain spectroscopy setup with an ErAs:InGaAs-based photoconductive dipole antenna used as the terahertz detector (*53*). The detector current is fed to a transimpedance amplifier (FEMTO DHPCA amplifier with an amplifier gain of $10^6$ V/A and a bandwidth of 1.8 MHz) followed by a lock-in amplifier (Zurich Instruments MFLI). The impact of the optical power on the generated terahertz spectrum is investigated for the nanoantenna array with a 2 μm nanoantenna length, 2 μm ground line width, and 0.5 μm gap between the nanoantenna array rows, fabricated on a 100-nm-thick undoped InAs layer grown on a 500-nm-thick p-type ($10^{19}$ cm$^{-3}$) InAs epilayer on a semi-insulating GaAs substrate. As shown in Fig. S1 and Fig. 2D, the generated terahertz power level has a linear dependence on the optical power level. This linear dependence is due to the high peak power of the femtosecond optical pulses, which result in an excessive number of free carriers in the InAs layer, screening the built-in electric field. The impact of the field screening becomes more apparent when observing the dependence of the generated terahertz power level on the optical power level in a continuous-wave (CW) operation mode with much lower peak optical power levels than those used in the pulsed operation mode. When the nanoantenna array is excited by two CW lasers with a terahertz frequency difference, the generated terahertz power shows a quadratic dependence on the optical power level (Fig. S11B inset). Up to a 110 dB dynamic range and radiation up to 5 THz is achieved at a 900 mW optical power when using 1550 nm optical pulses with a 120 fs pulsewidth and a 76 MHz repetition rate (Fig. S1). The terahertz radiation bandwidth can be further extended by using optical pulses with narrower pulsewidths. As shown in Fig. S2, a terahertz radiation bandwidth exceeding 6.5 THz is measured when using 1560 nm optical pulses with a 23 fs pulsewidth, 100 mW average power, and a 40 MHz repetition rate (Er-doped fiber laser amplifier, TOPTICA Photonics FemtoFiber pro IRS-II).

**Supplementary Text**

Carrier Lifetime Measurements:

Optical pump terahertz probe measurements are performed to measure the carrier lifetime of the undoped InAs and p-type InAs ($p = 10^{19}$ cm$^{-3}$) layers grown on a semi-insulating GaAs substrate. Both measurements are performed on 1-μm-thick InAs layers. As shown in Fig. S3, the carrier lifetime of the p-type InAs layer is measured as 5 ps. The carrier lifetime of the undoped InAs layer is estimated to be much larger than 1 ns.

Impulse Response Calculation:

Impulse response current of the nanoantenna array fabricated on a 100-nm-thick undoped InAs layer grown on a 500-nm-thick p-type InAs ($p = 10^{19}$ cm$^{-3}$) layer grown on a semi-insulating GaAs substrate is calculated. A three-dimensional optical simulation is first performed using a finite-difference time-domain solver (Lumerical) on one nanoantenna unit cell under a TM-polarized 1550 nm optical excitation and the carrier generation profile, $G(x, y, z)$, is obtained from this simulation. Next, the continuity equation is solved for electron density, $n(x, y, z, t)$, under an impulse optical generation rate $G \cdot \delta(t)$:

$$\frac{\partial n}{\partial t} = \frac{1}{q} \nabla \cdot \vec{J}_n + G \cdot \delta(t) - \frac{n}{\tau} \tag{1}$$

where $\vec{J}_n$ is the electron current density in A/m$^2$, $\delta(t)$ is the Dirac-delta function, $\tau$ is the electron lifetime in seconds, and q is the electron charge in C. Next, Equation (1) is solved for both undoped InAs and p-type doped InAs regions. Following assumptions are made for electron transport in the undoped InAs region:

1. Electron current is dominated by the drift current in the x-direction.
2. Electrons are assumed to drift at the saturation velocity, i.e., $\vec{J}_n = q\vec{v}_e n$, $|v_e| = 10^5$ m/s (*54, 55*).
3. Since the carrier lifetime of the undoped InAs layer is much larger than the transit time of the photogenerated carriers inside the undoped InAs region that drift to the nanoantenna contact (~1 ps), a carrier lifetime of $\tau \to \infty$ is assumed in the undoped InAs region.

Under these assumptions, Equation (1) in the undoped InAs region is modified to:

$$\frac{\partial n}{\partial t} = v_e \frac{\partial n}{\partial x} + G \cdot \delta(t) \tag{2}$$

Following assumptions are made for electron transport in the p-doped InAs region:

1. Electron current is dominated by the diffusion current in the x-direction, i.e., $J_{n,x} = qD_n \frac{\partial n}{\partial x}$.
2. Diffusion constant of the p-type InAs layer is chosen as $D_n = 1.36 \times 10^{-4}$ m$^2$/s, based on the experimental results reported in (*56*). Brooks-Herring's model is used to account for the additional degradation in mobility due to the high doping density.
3. The carrier lifetime of the p-type InAs layer is chosen as 5 ps, based on the measurement results shown in Fig. S3.

Under these assumptions, Equation (1) in the p-type doped InAs region is modified to:

$$\frac{\partial n}{\partial t} = D_n \frac{\partial^2 n}{\partial x^2} + G \cdot \delta(t) - \frac{n}{\tau} \qquad (3)$$

Considering current continuity at the interface between the undoped and p-doped InAs layers, Equations (2) and (3) are solved to obtain $n(x, y, z, t)$. The time-evolution of the carrier density, $n$, is shown in the supplementary video S2.

Next, the density of the electron current injected to the nanoantenna at each surface point $(y, z)$ is calculated as:

$$J_{injected}(0, y, z, t)\hat{x} = qv_e n(0, y, z, t)\hat{x} \qquad (4)$$

To calculate the induced current on the nanoantennas, the nanoantenna area is first divided into discrete regions, $i$, with center locations of $(y_i, z_i)$ and dimensions of $\Delta y$ and $\Delta z$. The overall injected current at the $i^{th}$ discrete region is calculated as

$$\tilde{I}_{injected}(y_i, z_i, f) = \int_{y_i - \frac{\Delta y}{2}}^{y_i + \frac{\Delta y}{2}} \int_{z_i - \frac{\Delta z}{2}}^{z_i + \frac{\Delta z}{2}} \tilde{J}_{injected}(0, y, z, f) dy dz \qquad (5)$$

$\tilde{J}_{injected}$ is the Fourier transform of the calcuated time domain current, $J_{injected}$. The overall induced current on the nanoantennas is computed using a finite-element-method-based electromagnetic solver (ANSYS-HFSS). The calculated injected currents at all of the discrete regions, $\tilde{I}_{injected}(y_i, z_i, f)$, are included as multiple current sources across the nanoantenna area. One nanoantenna unit-cell is simulated and periodic boundary conditions are used in the y and z directions to account for the impact of the current injected to the adjacent nanoantennas while calculating the induced impulse response current on the nanoantenna, $\tilde{I}_{impulse}(y, z, f)\hat{z}$. Figs S4A and S4B show the impulse response current of the nanoantenna array and its Fourier transform amplitude, respectively.

To calculate the overall induced current on the nanoantennas, $\tilde{I}_{induced}(y, z, f)\hat{z}$, the obtained impulse response is convolved with the temporal profile of the femtosecond optical pulse. Figs 4 A-F show the overall induced current on the nanoantennas as a function of frequency for different nanoantenna lengths and the decomposition of the overall induced current to the contribution of the injected current from various spots along the antenna length. As the current injection position is moved away from the nanoantenna tip and the nanoantenna-ground line intersection, it splits into two current components in opposite directions and with an approximately equal magnitude. This is because a fraction of the current that is injected to the adjacent nanoantennas flows to the neighboring nanoantennas and induces a current in the opposite direction. To better illustrate this phenomenon, we compare the induced current on a nanoantenna under three different scenarios: (1) when no neighboring nanoantenna is excited with an injected current (Fig. S5B); (2) when only the two neighboring nanoantennas are excited with an equal injected current (Fig. S5C); and (3) when six neighboring nanoantennas are excited with an equal injected current (Fig. S5D). As illustrated in Fig. S5B, when no neighboring nanoantenna is excited, the current flow is mostly oriented toward the ground line due to the impedance asymmetry created by the ground line. However, as the adjacent nanoantennas are excited (Figs S5C-D) a portion of their currents flows to the nanoantenna and produces a current flow in the opposite direction with an approximately equal magnitude. Furthermore, there is a rapid drop in the induced current closely after the current injection location. This is because the photogenerated electrons increase the substrate conductivity, which results in a fraction of the injected current leaking to the substrate, as illustrated in Fig S5A.

## Radiation Power Calculation:

To calculate the radiated power from the nanoantenna array, the radiation field of a single nanoantenna, $E_i(r, \theta, \phi, f)$, is calculated first. Since the nanoantenna length is much smaller than the radiation wavelength and since most of the radiated power flows toward the semiconductor substrate, the vector potential can be written as (*52*):

$$A_z(r, f) = \frac{\mu_0}{4\pi} \frac{e^{-jkr}}{r} \int_0^{L_a} \int_0^{W_a} \tilde{J}_{induced}(y', z', f) dy' dz' = \frac{\mu_0}{4\pi} \frac{e^{-jkr}}{r} S(L_a, W_a) \qquad (6)$$

where $\tilde{J}_{induced}(y, z, f)$ is the induced surface current on the nanoantennas with an A/m unit, $L_a$ and $W_a$ are the length and width of each nanoantenna, respectively, $S(L_a, W_a)$ is the result of the integral, $\mu_0$ is the permeability of the free space, $k = 2\pi f/c$ is the free space wavenumber, $r = \sqrt{x^2 + y^2 + z^2}$ is the observation point distance.

From the vector potential expression, the far-field radiated electric field for one nanoantenna can be written as (*52*):

$$E_i(r, \theta, \phi, f) \approx j\eta_0 k \sin\theta \frac{e^{-jkr}}{4\pi r} S(L_a, W_a)\hat{\theta} \qquad (7)$$

where $\eta_0$ is the wave impedance of the free space. Next, the radiation pattern of the nanoantenna array is calculated using the radiation pattern of one nanoantenna, $E_i(r, \theta, \phi, f)$ (*52*):

$$E_{array}(r, \theta, \phi, f) = E_i(r, \theta, \phi, f) AF_y(\theta, \phi, f) AF_z(\theta, f) \qquad (8)$$

where $AF_y$ and $AF_z$ are the array factor of the nanoantenna array in the y and z directions:

$$AF_y(\theta, \phi, f) = \frac{\sin\left(\frac{N_y \psi_y}{2}\right)}{\sin\left(\frac{\psi_y}{2}\right)}, \psi_y = p_y k \sin\phi \sin\theta \qquad (9)$$

$$AF_z(\theta, f) = \frac{\sin\left(\frac{N_z \psi_z}{2}\right)}{\sin\left(\frac{\psi_z}{2}\right)}, \psi_z = p_z k \cos\theta \qquad (10)$$

where $p_y$ and $p_z$ are the periodicity of the nanoantennas in the y and z directions, respectively, $N_y$ and $N_z$ are the number of nanoantennas in the y and z directions, respectively, and $k = 2\pi nf/c$ is the wavenumber at a given radiation wavelength ($n = n_{GaAs} = 3.6$). Finally, the radiated power from the nanoantenna array that propagates toward the semiconductor substrate is calculated as:

$$P(f) = \frac{1}{2\eta_0} \int_{-\pi/2}^{\pi/2} \int_0^{\pi} \left| E_{array}(\theta, \phi, f) \right|^2 r^2 \sin\theta d\theta d\phi \qquad (11)$$

## Surface Plasmon Excitation:

Periodicity of the nanoantennas in the y-direction is chosen as 440 nm to provide the necessary momentum to couple the photo-excited surface plasmon waves to the interface between the nanoantennas and the InAs substrate when excited by a TM-polarized optical beam at a 1550 nm wavelength (Fig. S7A). A 240-nm-thick $Si_3N_4$ anti-reflection coating, a 360-nm-thick nanoantenna width, and a 3/97-nm-thick Ti/Au nanoantenna height are used to increase the coupling efficiency of surface plasmon waves. To illustrate the impact of the excited surface plasmon waves, an alternative nanoantenna geometry is analyzed, which has a periodicity of 160 nm in the y-direction. The momentum provided by this periodicity is larger than the momentum required for the excitation of surface plasmon waves (Fig. S7B). Although the optical transmission to the InAs layer provided by this nanoantenna (Fig. S7C) is much higher than that of the plasmonic nanoantenna, the plasmonic nanoantenna provides 7 times higher optical absorption within a 100 nm depth inside the InAs layer, where the built-in electric field strength is maximized (Fig. S7D).

To quantify the overlap between the optical absorption and built-in electric field, an overlap integral function, OI(x), is defined as:

$$OI(x) = \frac{\int_0^x \int_0^{w_a} E_{bi}(y', x') I_{abs}(x', y') dy' dx'}{\int_0^{\Lambda} I_{in}(y') dy'} \tag{12}$$

where $E_{bi}$ is the built-in electric field shown in Fig. 2E color plots, $I_{abs}$ is the absorbed optical beam intensity at a 1550 nm wavelength (Figs. S7 A and B insets), $I_{in}$ is the incident optical beam intensity on the nanoantenna array, and x is the depth in the InAs substrate. As expected, the calculated overlap integral values shown in Fig. S7E, are consistently higher for the nanoantenna design that supports surface plasmon waves.

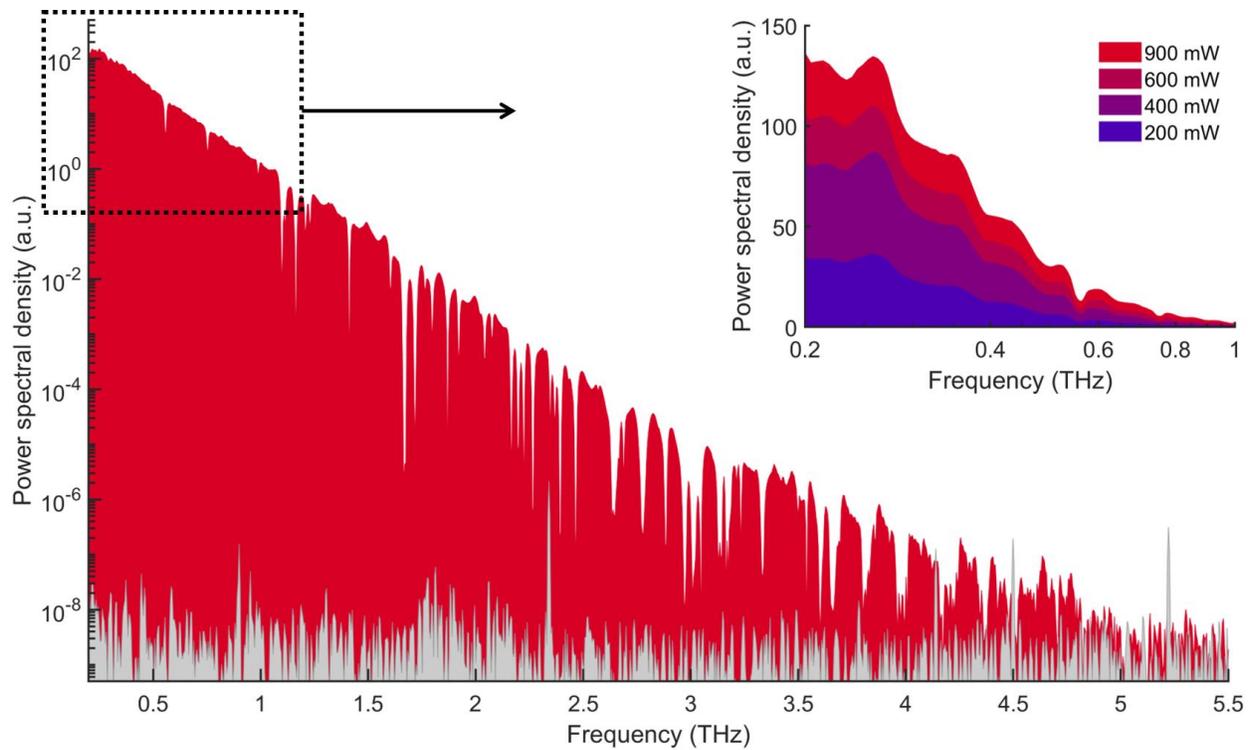

**Fig. S1.** The measured terahertz radiation spectrum (shown in red) along with the noise spectrum (shown in gray) when the nanoantenna array is excited by optical pulses with a 1550 nm center wavelength, 900 mW power, 120 fs pulsewidth, and 76 MHz repetition rate. 650 time-domain traces are captured and averaged to resolve this terahertz spectrum. Dependence of the radiation spectrum on the optical power level is shown in the inset.

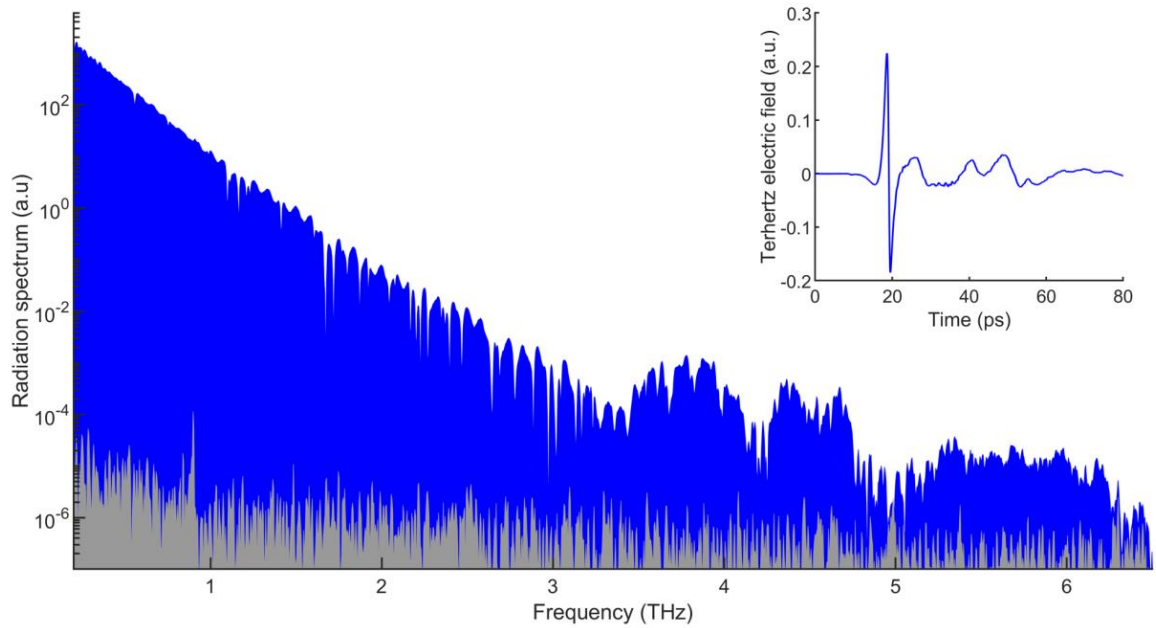

**Fig. S2.** The measured terahertz radiation spectrum (shown in blue) along with the noise spectrum (shown in gray) when the nanoantenna array is excited by optical pulses with a 1560 nm center wavelength, 100 mW power, and 23 fs pulsewidth. 1000 time-domain traces are captured and averaged to resolve this terahertz spectrum. The time-domain radiated terahertz pulse is shown in the inset.

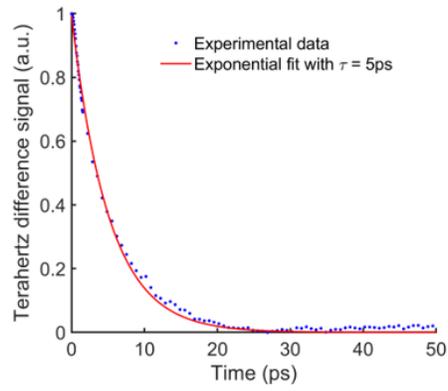

**Fig. S3.** Optical pump terahertz probe measurement of the p-type InAs (p = 10$^{19}$ cm$^{-3}$) layer grown on a semi-insulating GaAs substrate.

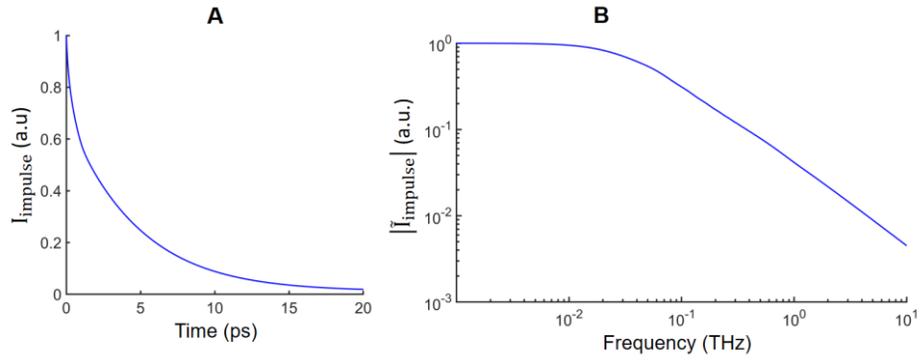

**Fig. S4.** Impulse response current of the nanoantenna array and its Fourier transform amplitude are shown in (**A**) and (**B**), respectively.

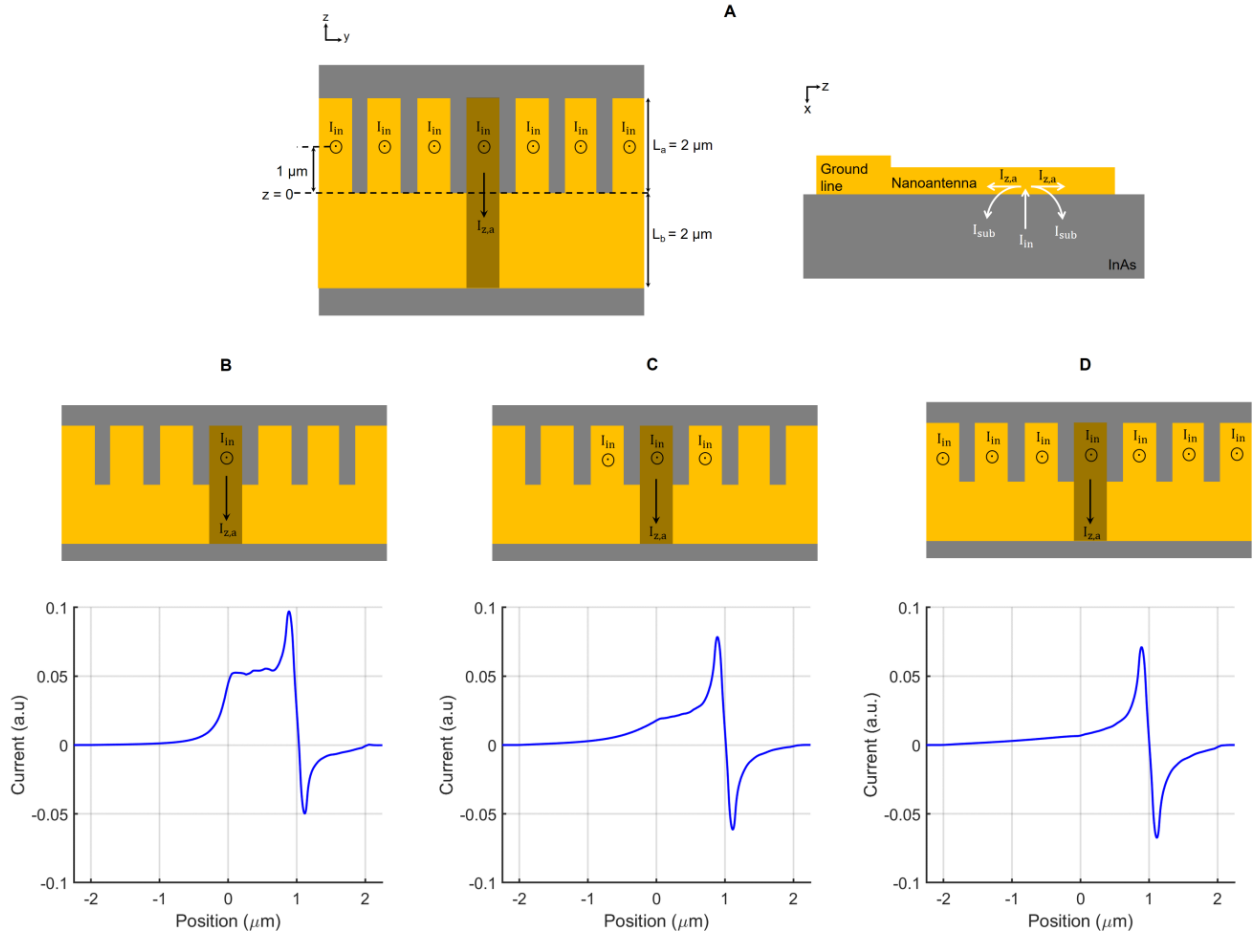

**Fig. S5.** (**A**) Top view and side view of the simulated nanoantenna array to investigate the induced current on each nanoantenna. The induced current on the shaded nanoantenna when (**B**) the neighboring nanoantennas are not excited, (**C**) the two neighboring nanoantennas are excited, and (**D**) six neighboring nanoantennas are excited.

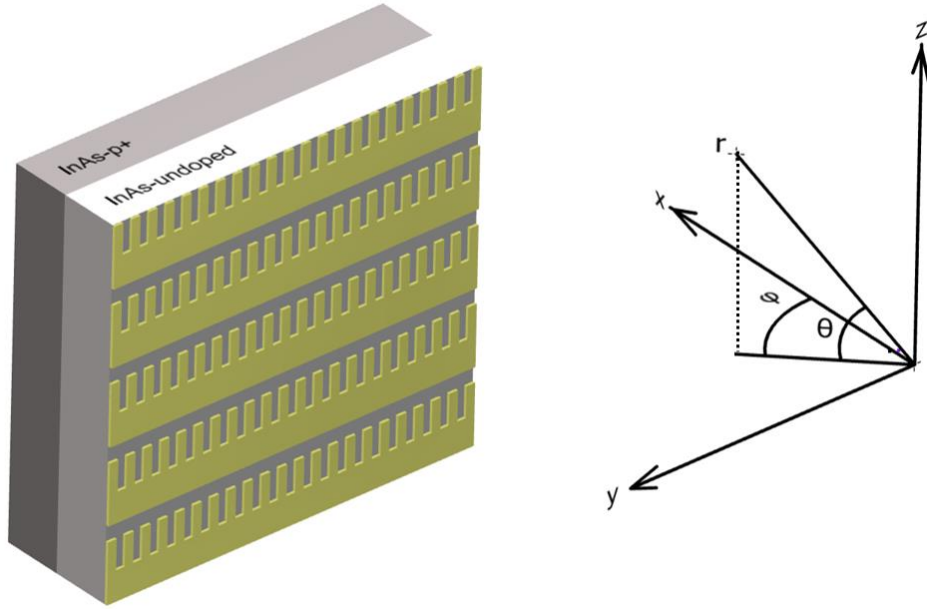

**Fig. S6.** Coordinate axis used for the radiation power calculations.

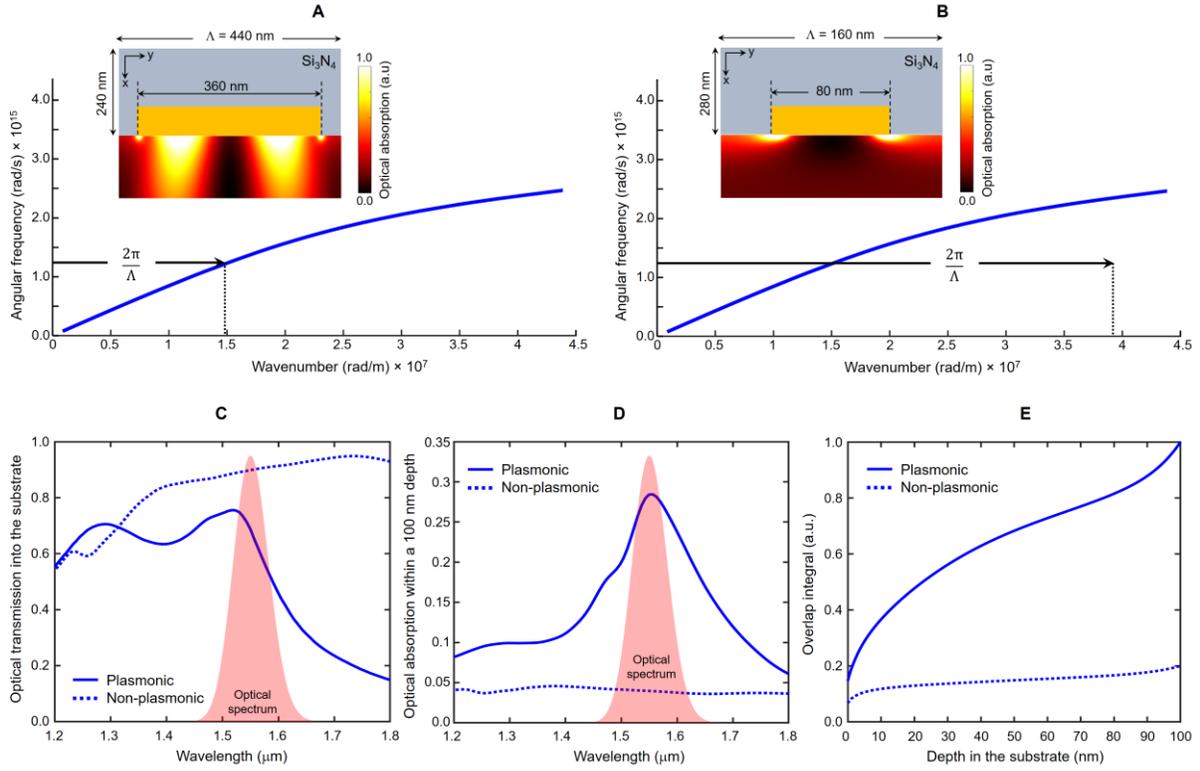

**Fig. S7.** (**A**) Surface plasmon dispersion curve and the momentum provided by nanoantennas with a periodicity of Λ = 440 nm. Optical absorption profile inside the InAs layer when a 1550 nm TM-polarized optical excitation is incident on the nanoantennas (inset). (**B**) Surface plasmon dispersion curve and the momentum provided by nanoantennas with a periodicity of Λ = 160 nm. Optical absorption profile inside the InAs layer when a 1550 nm TM-polarized optical excitation is incident on the nanoantennas (inset). Optical transmission spectrum to the InAs layer, optical absorption spectrum within a 100 nm depth in the InAs layer, and the overlap integral calculated as a function of depth in the substrate for both nanoantenna designs are shown in (**C**), (**D**), and (**E**) respectively. All of the optical simulations are performed using a finite-difference time-domain solver (Lumerical).

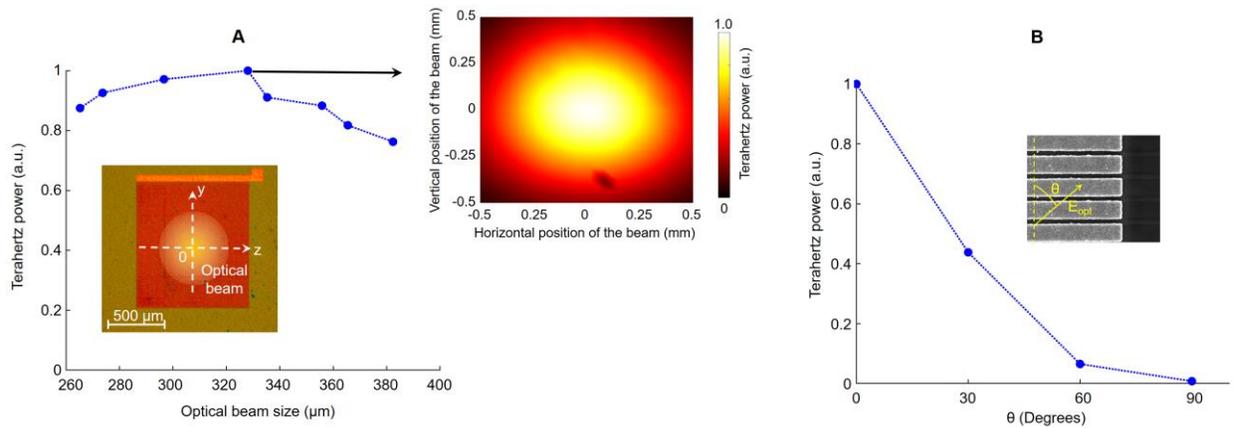

**Fig. S8.** Optical beam size, position, and polarization requirements. (**A**) The measured terahertz radiation power from the nanoantenna array for different optical beam sizes and beam positions. Beam size is defined as the beam diameter at which the optical intensity drops to $1/e^2$ (13.5%) of its peak value. The measurement results show less than a 25% reduction in the radiation power when the optical beam size is deviated from the optimum value (330 μm) by 50 μm. When the optical beam size is reduced relative to this optimum value, the wavelength conversion efficiency is reduced due to the carrier screening effect. When the optical beam size is increased relative to this optimum value, the wavelength conversion efficiency is reduced due to the destructive interference of the radiation from the relatively distant nanoantennas. The results also show less than a 25% reduction in the radiation power when the optical beam position is deviated from the optimum position (center of the nanoantenna array) by 250 μm. Therefore, wavelength conversion through plasmon-coupled surface states has a high tolerance for optical beam size variations and misalignment. (**B**) The measured terahertz radiation power from the nanoantenna array as a function of optical polarization. The wavelength conversion efficiency drops when deviating from the optimum polarization for the excitation of surface plasmons (orthogonal to the nanoantennas) due to the reduction in the number of the photogenerated electrons at the semiconductor surface.

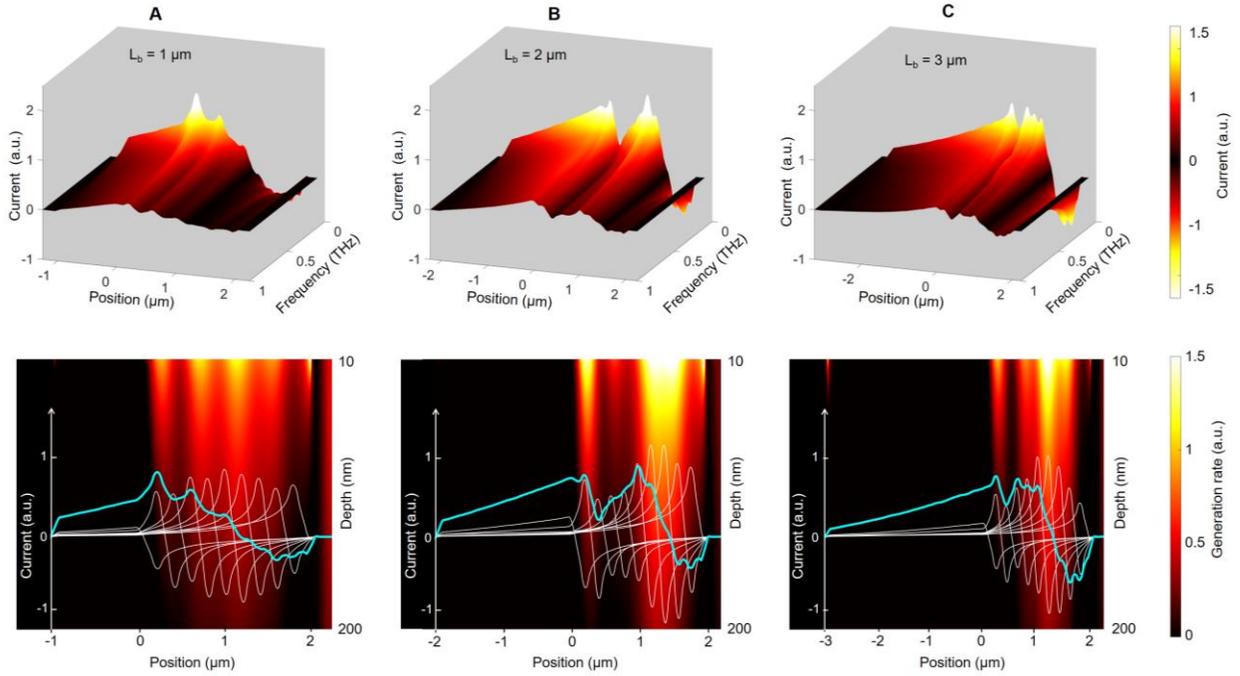

**Fig. S9.** (**A**)-(**C**) Top: The induced current on the nanoantennas as a function of frequency when the ground line width ($L_b$) is varied from 1 μm to 3 μm. The nanoantenna length ($L_a$) and the gap between the nanoantenna array rows ($L_g$) are chosen as 2 μm and 0.5 μm, respectively. The ground line junction with the nanoantenna is located at the 0 position and the nanoantenna is located between 0 and 2 μm positions along the z-axis. (**A**)-(**C**) Bottom: Decomposition of the total induced current on the nanoantennas (teal lines) to the individual contributions of the injected currents from different positions of the nanoantennas (white lines) at 0.2 THz. The background color maps show the electron generation profiles averaged over the nanoantenna width.

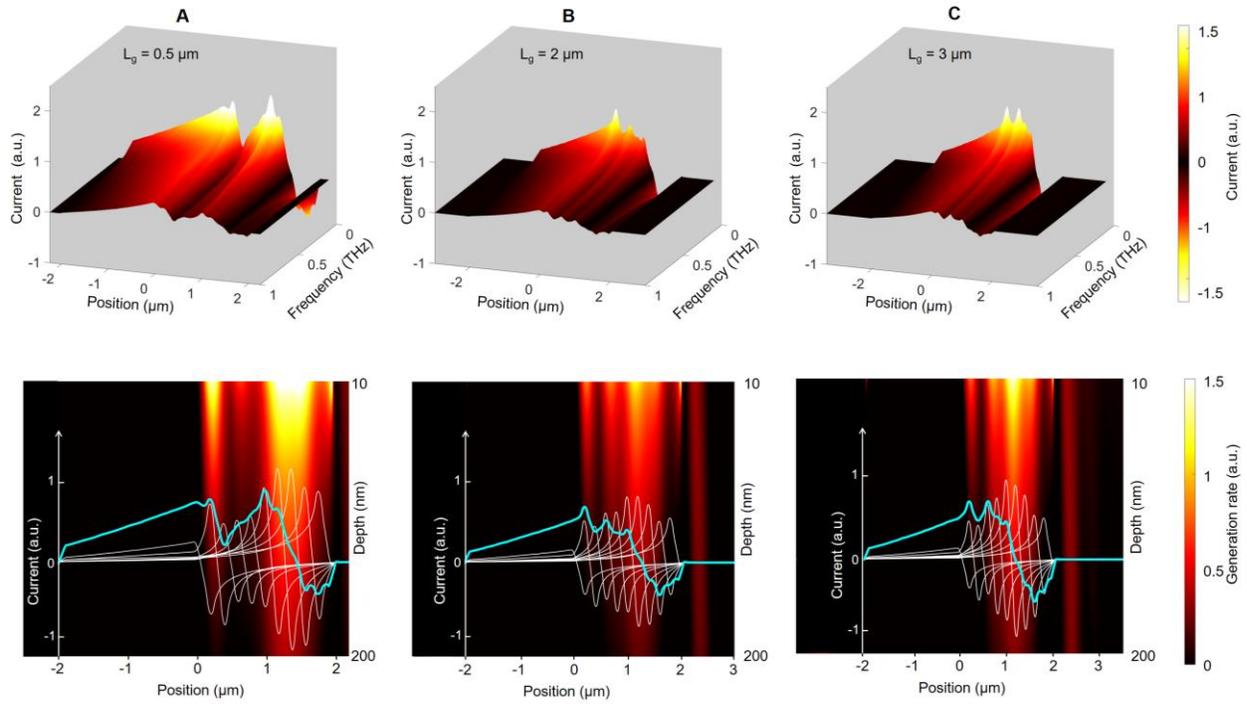

**Fig. S10.** (**A**)-(**C**) Top: The induced current on the nanoantennas as a function of frequency when the gap between the nanoantenna array rows ($L_g$) is varied from 0.5 μm to 3 μm. The nanoantenna length ($L_a$) and the ground line width ($L_b$) are chosen as 2 μm and 2 μm, respectively. The ground line is located between -2 μm and 0 positions and the nanoantenna is located between 0 and 2 μm positions along the z-axis. (**A**)-(**C**) Bottom: Decomposition of the total induced current on the nanoantennas (teal lines) to the individual contributions of the injected currents from different positions of the nanoantennas (white lines) at 0.2 THz. The background color maps show the electron generation profiles averaged over the nanoantenna width.

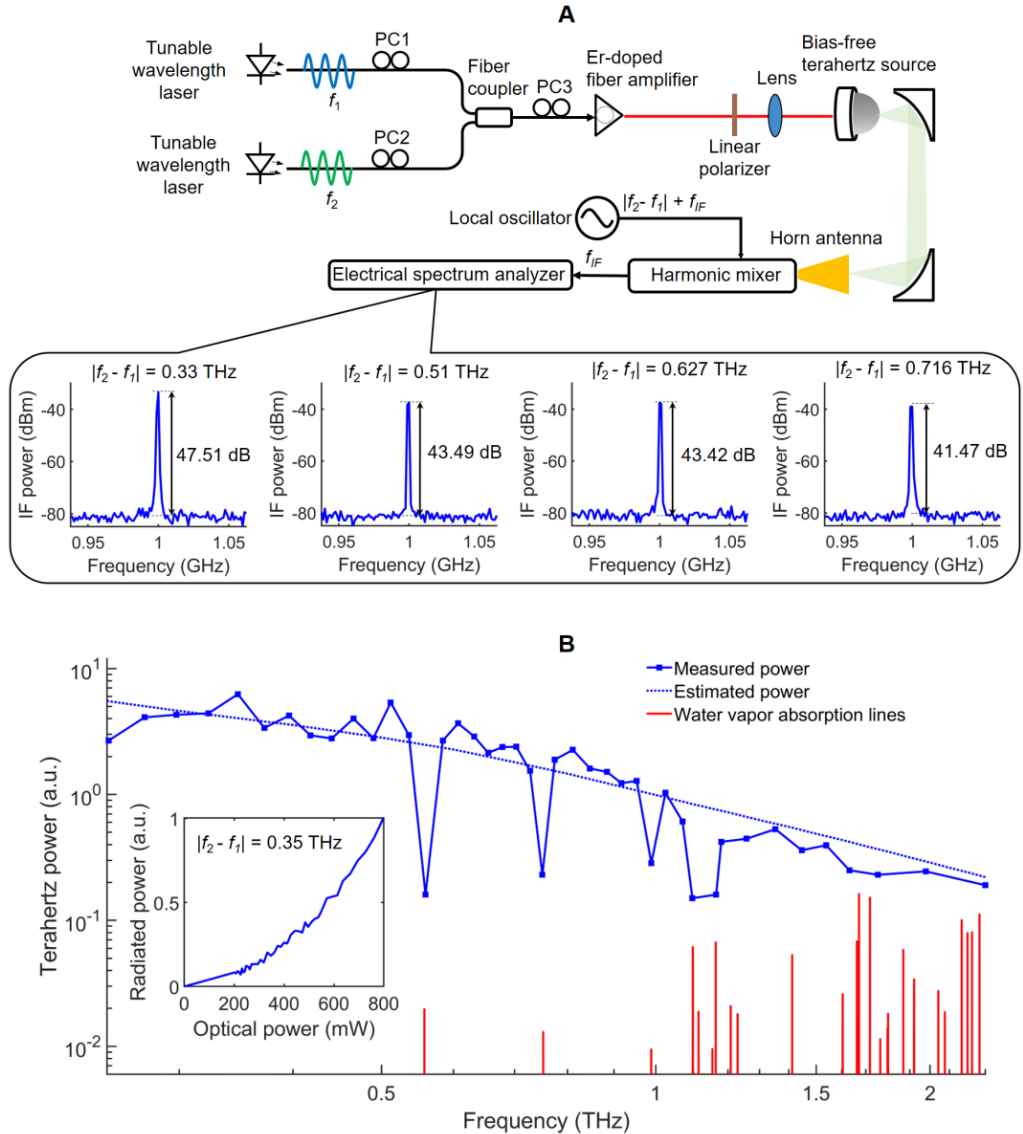

**Fig. S11.** (**A**) Top: Experimental setup used to measure CW radiation from the nanoantenna array. The nanoantenna array is excited by two 1550 nm CW lasers (Santec TSL-510 and New Focus TLB-6730-P) with equal optical power levels. The radiation frequency is tuned by adjusting the beat frequency of the two lasers. In this experimental setup, the beat frequency is varied between 0.23 THz and 2.3 THz. The generated radiation from the nanoantenna array is routed to a harmonic mixer (VDI MixAMC with WR 1.5 and WR 2.2 waveguides) to down-convert it to an intermediate frequency (IF) signal in the GHz frequency range. The IF signal for each optical beat frequency is measured using an electrical spectrum analyzer (HP8592L), as shown in the inset. (**B**) The measured radiation power from the nanoantenna array over the 0.23-2.3 THz frequency range using a pyroelectric detector. The observed power roll-off as a function of frequency agrees with the roll-off predicted by the calculated impulse response. The deviations at low frequencies are due to the poor coupling of the radiation to the pyroelectric detector. The dips in the spectrum match the water vapor absorption lines (*57*), shown in red. The inset shows the dependence of the radiated power on the incident optical power at 0.35 THz. Unlike the pulsed operation mode, the radiated power follows a quadratic relation with the incident optical power.

**Movie S1.**
Illustration of free carrier generation and transport dynamics inside the InAs lattice, where a nanoantenna array couples photo-excited surface plasmons to the surface states.

**Movie S2.**
Time-evolution of the photo-generated carrier density inside the InAs layer.